\documentclass[journal=jacsat,manuscript=articlee]{achemso}
\pdfoutput=1

\usepackage{xcolor} 
\definecolor{ocre}{RGB}{0,102,25}
\usepackage[version=3]{mhchem} 
\usepackage[T1]{fontenc}       
\usepackage{fancyhdr}
\usepackage{graphicx}
\usepackage[]{algorithm2e}
\usepackage{amsmath}
\usepackage{caption}
\usepackage{setspace}
\usepackage[citecolor=blue, urlcolor=ocre, citebordercolor=ocre, urlbordercolor=ocre, linkbordercolor=ocre]{hyperref}
\usepackage{comment}

\SectionNumbersOn

\author{Andrew Abi Mansour}
\affiliation{Department of Chemistry and
Center for Theoretical and Computational Nanoscience, \\
Indiana University , Bloomington, Indiana 47405}

\author{Peter J. Ortoleva}
\email{ortoleva@indiana.edu}

\affiliation{Department of Chemistry and
Center for Theoretical and Computational Nanoscience, \\
Indiana University , Bloomington, Indiana 47405}

\title{Reverse Coarse-graining for Equation-free Modeling: Application to Multiscale Molecular Dynamics}


\begin{document}

\begin{abstract}
 Constructing atom-resolved states from low-resolution data is of practical importance in many areas of science and engineering.
 This problem is addressed in this paper in the context of multiscale factorization methods for molecular dynamics. These methods capture the crosstalk
 between atomic and coarse-grained scales arising in macromolecular systems. This crosstalk is accounted for by Trotter factorization,
 which is used to separate the all-atom from the coarse-grained phases of the computation. In this approach, short molecular dynamics runs are used to advance in time the coarse-grained variables, which in turn guide the all-atom state. To achieve this coevolution, an all-atom microstate consistent with the updated coarse-grained variables must be recovered. This recovery is cast here as a non-linear optimization problem that is solved with a quasi-Newton method. The approach yields a Boltzmann-relevant microstate whose coarse-grained representation and some of its fine-scale features are preserved. Embedding this algorithm in multiscale factorization is shown to be accurate and scalable for simulating proteins and their assemblies.
\end{abstract}

\maketitle

\section{Introduction}
Mesoscopic systems such as nanocapsules, viruses, and ribosomes evolve through the coupling of processes across multiple scales in space and time.
Therefore, a theory of the dynamics of these systems must somehow account for the coevolution of coarse-grained (CG)
and microscropic (atomistic) variables. Multiscale coevolution
\cite{Ortoleva2005,Ortoleva2008,Ortoleva2009,Cheluvaraja2010,abi2016implicit}, an equation-free method \cite{Kevrekidis2000, Gear2002, Gear2003, Kevrekidis2009}, has been proposed as an alternative to purely coarse-grained methods
\cite{Bahar1997, Voth2008, Reith2003, Schulten2006, Muller2002, Broughton1998};
coevolution methods operate via a cycle consisting of microscopic and coarse-grained phases. These methods do not involve deriving CG dynamical equations in closed form. Instead, the CG dynamics follow directly from the microscopic dynamics. In contrast,
traditional CG methods evolve large-scale structural variables via phenomenological equations \cite{Bahar1997, Voth2008, Schulten2006},
and they do not provide information on the evolving microstate(s).

\begin{figure}[h]
\centering
\includegraphics[scale=0.5]{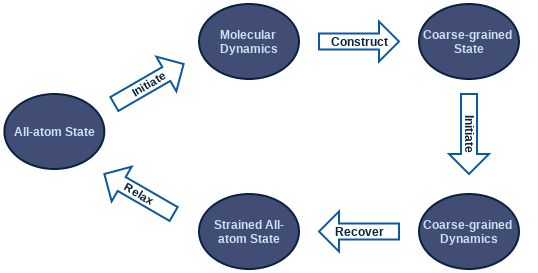}
\caption{In multiscale coevolution methods, a molecular dynamics run is initiated to collect information needed to advance the coarse-grained state in time. Afterwards, the all-atom state is recovered to begin another dual-phase step.}
\label{fig:mcf}
\end{figure}

The focus of this study is multiscale factorization \cite{AbiMansour2014,Sereda2014}, an equation-free coevolution method applied to molecular dynamics (MD) for macromolecular systems such as proteins and their assemblies. A necessary condition for an efficient multiscale simulation is the separation of timescales between the atomistic fluctuations and coherent, slow changes captured by the CG variables \cite{Ortoleva2008,Ortoleva2009,Cheluvaraja2010,Singharoy2011,Singharoy2012}. Furthermore, the MD phase of the multiscale computation should be sufficiently long to generate a representative ensemble of fluctuations in the CG momenta (i.e. longer than the `stationarity time' \cite{AbiMansour2014}). To complete the multiscale cycle, a microstate  consistent with the updated CG variables must be constructed before the MD phase of the computation is resumed. This multiscale approach is summarized in the flowchart shown in Fig. (\ref{fig:mcf}). In this way, the entire multiscale computation follows directly from an interatomic force field, and avoids the need for introducing phenomenological CG governing equations and the uncertainty associated with them. However, recovering a microstate consistent with the CG variables is an ill-posed problem \cite{hansen1998rank,Ensing2010} because there is an information gap between the CG and fine-grained (FG) descriptions. This reverse coarse-graining is cast as a non-linear optimization problem that is solved with a quasi-Newton method that leads to sparse matrices, thus enabling an efficient and scalable fine-graining algorithm that is hereafter refered to as `microstate sparse reconstruction' (MSR). In this paper, it is shown how embedding MSR in MF yields an accurate and scalable method for simulating macromolecular systems.

The mathematical framework for MSR is outlined in Sec. \ref{sec:theory}, and its implementation for distributed systems is described in Sec. \ref{sec:impl}. MSR is demonstrated for several macromolecular systems in Sec. \ref{sec:demo}, and conclusions are drawn in Sec. \ref{sec:conclusion}.

\section{Theory} \label{sec:theory}
For an efficient multiscale simulation, it is necessary to reduce the all-atom description represented by $N$ atoms to a set of $N_{CG}$ variables ($N_{CG} \ll N$)
that capture the coherent deformation of the system. These CG variables must be chosen such that they evolve on a timescale much greater than that of
fluctuating atoms. Let $\mathbf{\phi}$ denote a set of CG variables such that
\begin{equation}
\phi_{\alpha} = \mathbf{Q}(\mathbf{r}^0) \mathbf{r}_{\alpha} , \label{eq:dimred}
\end{equation}
where $\mathbf{r}_{\alpha}$ is a vector of all atomic positions with $\alpha$ corresponding to the $x$, $y$, or $z$ axis, and $\mathbf{Q}$ is a matrix of dimensions $N_{CG} \times N$ that depends on the atomic positions of a reference configuration denoted $\mathbf{r}^0$. Initially, the reference structure introduces a
configuration determined by data collected from x-ray, cryo-EM, or other experimental techniques. However, at later times, the reference structure is taken from a previous time step in a discrete time evolution sequence. Thus, the CG variables specify how the structure is deformed
from this reference configuration in the course of the simulation. While coarse-graining can be uniquely defined for a system, the inverse problem of finding a microstate from a given CG description is ill-posed and therefore has no unique solution. This problem is addressed below.

\subsection{Microstate reconstruction}
\begin{figure}[h]
\centering
	\includegraphics[scale=0.3]{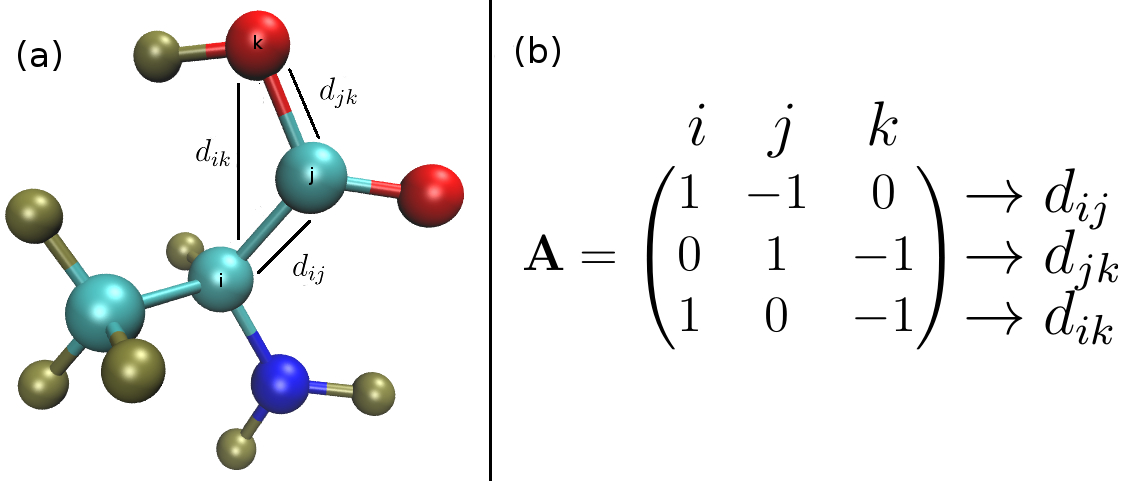}
	\caption{The adjacency-like matrix $\mathbf{A}$ (b) for the three atoms labeled $i$, $j$, and $k$ shown in (a). To preserve the two bond lengths and harmonic angle of the triplet of atoms ($i-j-k$), the three interatomic distances $d_{ij}$, $d_{jk}$, and $d_{ik}$ are constrained to their equilibrium values extracted from MD.}
	\label{fig:graph}
\end{figure}

A challenge in multiscale equation-free methods is recovering an all-atom configuration consistent with the updated CG variables. This is formulated here as an optimization problem that minimizes the norm of the difference between $\mathbf{r}_{\alpha}$ and $\mathbf{r}^0_{\alpha}$ over all three cartesian components,
subject to the constraints imposed by the updated CG variables (Eq. (\ref{eq:dimred})). Thus, the CG constraints act as a perturbation that guides the all-atom microstate to a configuration which minimizes deviation from the reference configuration and is consistent with the imposed CG description. To take microstate effects
such as those imposed by stiff bonds into account, FG constraints are included in order to enforce constant bond lengths and harmonic angles (Figure \ref{fig:graph}). This preserves key aspects of the microstructure.

The optimization problem is formulated in terms of a quadratic function as follows:
\begin{equation}
\min_{\mathbf{r}_{\alpha}} f\left(  \mathbf{r} \right) = \frac{1}{2} \sum_{\alpha'} \left( \mathbf{r}_{\alpha'} - \mathbf{r}_{\alpha'}^0 \right)^T
\left( \mathbf{r}_{\alpha'} - \mathbf{r}^0_{\alpha'} \right),
\end{equation}
subject to the following constraints:
\begin{align}
\phi_{\alpha} - \mathbf{Q}(\mathbf{r}_{\alpha}^0) \mathbf{r}_{\alpha} &= \mathbf{0}, \label{eq:cons_CG} \\
\sum_{\alpha} D(\mathbf{A}\mathbf{r}_{\alpha}) (\mathbf{A}\mathbf{r}_{\alpha}) - D(\mathbf{l}_{\alpha}) \mathbf{l}_{\alpha} &= \mathbf{0}, \label{eq:cons_FS}
\end{align}

where $D(\mathbf{v})$ denotes a diagonal matrix whose entries are equal to those of vector $\mathbf{v}$, $\mathbf{l}_{\alpha}$ represents a vector of interatomic distances (for the atomic bonds and harmonic angles) computed from the MD phase for each $\alpha$, and $\mathbf{A}$ is an adjacency-like matrix that captures the location of each atomic index in every equation of the FG constraints, i.e., for the $k^{th}$ constraint spanning atoms $i$ and $j$, row $k$ in $\mathbf{A}$ has $+1$ entry at column $i$, and $-1$ at column $j$, while all remaining columns have zero entries (Figure \ref{fig:graph}). For convenience, the inter-atomic distance squared was used in the equations of the FG constraints. 

The Lagrangian $\mathcal{L}$ of the above optimization problem is
\begin{equation}
\mathcal{L}(\mathbf{r}, \mathbf{\mu}, \mathbf{\lambda}) = f\left(\mathbf{r} \right) + \sum_{\alpha} \mu_{\alpha }^T \left( \phi_{\alpha} - \mathbf{Q}(\mathbf{r}_{\alpha}^0) \mathbf{r}_{\alpha} \right)
+ \lambda^T \sum_{\alpha} \left( D(\mathbf{A}\mathbf{r}_{\alpha}) (\mathbf{A}\mathbf{r}_{\alpha}) - D(\mathbf{l}_{\alpha}) \mathbf{l}_{\alpha} \right)  \label{eq:optimization}
\end{equation}
The Lagrange multipliers represented by the vector $\mathbf{\mu}_{\alpha}$ ensure
the recovered microstate is consistent with the updated CG state for every $\alpha$ direction, and those
represented by the vector $\mathbf{\lambda}$ enforce constant bond lengths and harmonic angles. While the Lagrangian does not necessarily admit a unique minimum, this does not contradict the physics of the problem since there is an ensemble of microstates consistent with a given CG description. The numerical scheme for minimizing the Lagrangian in Eq. (\ref{eq:optimization}) is covered in the next section. 

\section{Implementation} \label{sec:impl}

MSR was implemented using ProtoMD \cite{protoMD}, a prototyping toolkit written in python for multiscale MD. The paralellization of the algorithm was done with the aid of PETSc \cite{PETSc-efficient,PETSc-user-ref,PETSc-web-page} (Portable, Extensible Toolkit for Scientific computation) while SWIG \cite{SWIG} (Simplified Wrapper and Interface Generator) was used to interface the C++ modules (that use PETSc) with the python code (ProtoMD source code). The MSR code is freely available on github \cite{MSR}. The VdW forces were modeled using a Lennard-Jones potential that is slightly modified by shifting the interatomic distance by $1$ \AA\ to prevent the repulsive term from diverging to infinity when this distance approaches $0$.

\subsection{Coarse-graining}
The space-warping method is used as a dimensionality reduction technique \cite{Khuloud2002,Joshi2012} in this study because the CG variables obtained with this
method are slowly varying in time for the systems considered here. In this method, the mapping matrix $\mathbf{Q}$ in Eq. (\ref{eq:dimred}) is constructed from a set of
products of three Legendre polynomials which are functions of the $x$, $y$, and $z$ positions of the reference configuration of orders $k_x$, $k_y$, and $k_z$, respectively. The total order of the method is $k_m$ such that $k_m \geq k_x + k_y +k_z$. A brief review of the space-warping method and the particular form of $\mathbf{Q}$ used here is covered in Appendix \ref{app:swm}.

\subsection{Regularizing the Lagrangian}

Using Newton's method to minimize Eq. (\ref{eq:optimization}) is not possible because the Hessian of the Lagrangian is ill-conditioned \cite{Optimization2006,elden1977algorithms}. Instead, an $L_2$ regularization (Appendix \ref{app:regular}) is imposed to obtain an approximate numerical solution. First, the Lagrangian in Eq. (\ref{eq:optimization}) is recast in the form
\begin{equation}
\mathcal{L}(\mathbf{r}) = f\left(\mathbf{r}\right) + \sum_{\alpha} \mu_{\alpha }^T \left( \phi_{\alpha} - \mathbf{Q}(\mathbf{r}_{\alpha}^0) \mathbf{r}_{\alpha} \right)
+ \lambda^T \sum_{\alpha'} \left( D(\mathbf{A}\mathbf{r}_{\alpha'}) (\mathbf{A}\mathbf{r}_{\alpha'}) - D(\mathbf{l}_{\alpha'}) \mathbf{l}_{\alpha'} \right) + \beta^2 \lambda ^T \lambda, \label{eq:optimization2}
\end{equation}
where $\beta$ is a regularization parameter set to $1$, and the penalty term $\lambda ^T \lambda$ keeps the norm of $\lambda$ to a minimum. The Lagrangian of Eq. (\ref{eq:optimization2}) is minimized by setting its gradient to zero with respect to the atomic positions and Lagrange multipliers. This yields
\begin{align}
\mathbf{r}_{\alpha} &= \mathbf{r}_{\alpha}^0 - \mathbf{J}_{\mathbf{r}_{\alpha}}^T \mathbf{\lambda} + \mathbf{Q}^T \mathbf{\mu}_{\alpha}, \label{eq:coords_update}\\
\mathbf{\phi}_{\alpha} &= \mathbf{Q} \mathbf{r}_{\alpha} \label{eq:cons_cg} \\
\sum_{{\alpha}} D(\mathbf{A}\mathbf{r}_{\alpha}) (\mathbf{A}\mathbf{r}_{\alpha}) &= \sum_{{\alpha}} D(\mathbf{l}_{\alpha}) \mathbf{l}_{\alpha} - \beta^2 \lambda, \label{eq:cons_dist}
\end{align}
where $\mathbf{J}_{\mathbf{r}_{\alpha}}$ is the Jacobian of Eq. (\ref{eq:cons_dist}) with respect to the atomic positions, and it is given by
\begin{equation}
\mathbf{J}_{\mathbf{r}_{\alpha}} = 2 \times D(\mathbf{A}\mathbf{r}_{\alpha}) \mathbf{A}.
\end{equation}
Equations (\ref{eq:coords_update}-\ref{eq:cons_dist}) are solved with a quasi-Newton method in a way analogous to MD constraint algorithms \cite{Barth2004}. This is achieved by decoupling the Lagrange multipliers from the atomic positions. First, Eq. (\ref{eq:coords_update}) is recast in terms of the unconstrained atomic positions vector, $\mathbf{r}^u_{\alpha}$ (which is initially set to $\mathbf{r}^0_{\alpha}$) such that
\begin{equation}
\mathbf{r}_{\alpha} = \mathbf{r}_{\alpha}^u - \mathbf{J}_{\mathbf{r}_{\alpha}^u}^T \mathbf{\lambda} + \mathbf{Q}^T \mathbf{\mu}. \label{eq:coords_update2}
\end{equation}
The Jacobian $\mathbf{J}_{\mathbf{r}_{\alpha}^u}$ is evaluated at $\mathbf{r}_{\alpha} = \mathbf{r}^u_{\alpha}$. The initial guess in every Newton iteration for the Lagrange multipliers is therefore always zero. The Lagrange multipliers are then decoupled and updated separately via
\begin{align}
\mathbf{J}_{\mathbf{\lambda}} \mathbf{\lambda} &= \sum_{\alpha} D(\mathbf{l}_{\alpha}) \mathbf{l}_{\alpha} - D(\mathbf{A}\mathbf{r}_{\alpha}^u) (\mathbf{A}\mathbf{r}_{\alpha}^u), \label{eq:linear1}\\
\mathbf{Q}\mathbf{Q}^T \mathbf{\mu}_{\alpha} &= \mathbf{\phi}_{\alpha} - \mathbf{Q} \mathbf{r}_{\alpha}^u. \label{eq:linear2}
\end{align}
Using the chain rule, the Jacobian $\mathbf{J}_{\mathbf{\lambda}}$, evaluated at $\lambda = 0$, is found to be
\begin{equation}
\mathbf{J}_{\mathbf{\lambda}} |_{\mathbf{\lambda}=\mathbf{0}} = \sum_{\alpha} \mathbf{J}_{\mathbf{r}_{\alpha}^u} \mathbf{J}^T_{\mathbf{r}_{\alpha}^u} + \beta^2 \mathbf{I}.
\end{equation}
Once Eqs. (\ref{eq:linear1}-\ref{eq:linear2}) are solved, the Lagrange multipliers are used in Eq. (\ref{eq:coords_update2}) to update
the atomic positions vector, $\mathbf{r}_{\alpha}$. The unconstrained positions vector $\mathbf{r}^u_{\alpha}$ is then equated to $\mathbf{r}_{\alpha}$,
the Lagrange multipliers set to zero, and the procedure is repeated until the maximum atomic displacement is below $10^{-2}$ \AA. Algorithm \ref{algo:msr} summarizes
this.
\begin{algorithm}
\While{$error \geq tol$} {
	\emph{Construct $\mathbf{J}_{\mathbf{r}_{\alpha}^u}$ for each $\alpha$}\;
	\emph{Assemble $\mathbf{J}_{\mathbf{\lambda}}$}\;
	\emph{Compute $\lambda$ and $\mu$ by solving Eqs. (\ref{eq:linear1} - \ref{eq:linear2})}\;
	\emph{Update atomic positions via Eq. (\ref{eq:coords_update2})}\;
	\emph{$\lambda \leftarrow 0 $}\;
	\emph{$\mu \leftarrow 0 $}\;
	\emph{Compute error}\;
	}
 \caption{MSR updates the atomic positions and Lagrange multipliers in an alternating way such that both the CG and FG constraints are satisfied
 to within a certain tolerance.}
 \label{algo:msr}
\end{algorithm} 

\subsection{Sparse storage}
\begin{figure}[H]
\centering
\includegraphics[scale=0.4]{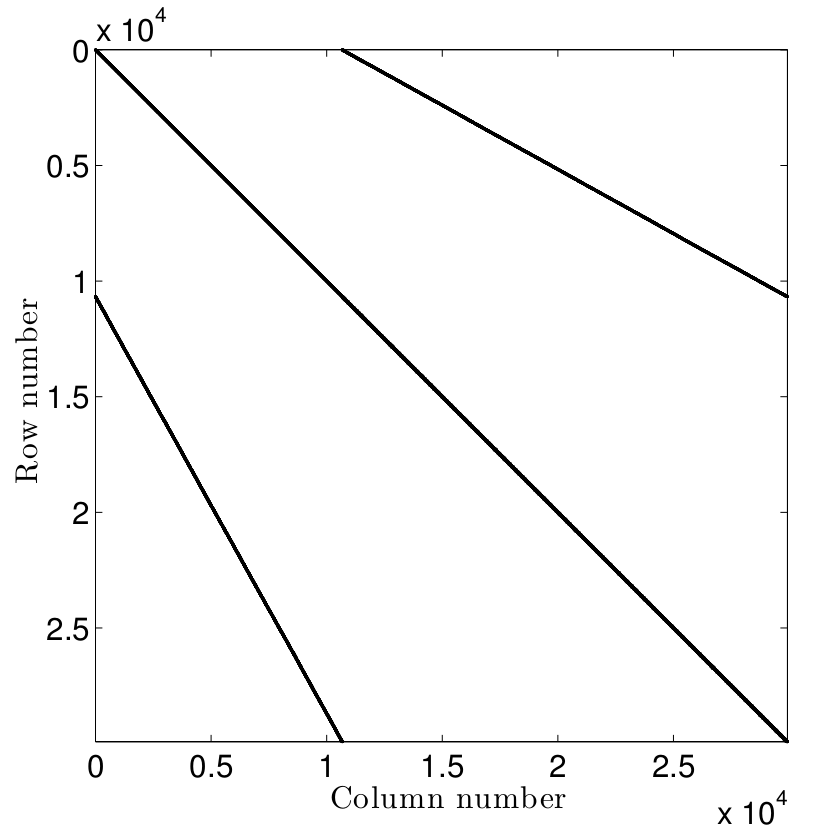}
\caption{The sparsity pattern of $\mathbf{J}_{\mathbf{\lambda}}$ for Lactoferrin.
In total, there are $29,923$ FG constraints, and $59,846$ non-zero entries, which makes $\mathbf{J}_{\mathbf{\lambda}}$ $99.99\%$ sparse.  }
\label{fig:sparse}
\end{figure}

The efficiency and scalability of MSR stem from the sparse structure of the Jacobians $\mathbf{J}_{\mathbf{r}_{\alpha}}$ and $\mathbf{J}_{\lambda}$.
Thus, these matrices are stored in compressed sparse row format \cite{Timothy2006} using PETSc \cite{PETSc-efficient,PETSc-user-ref,PETSc-web-page}. For example, the sparsity pattern for $\mathbf{J}_{\mathbf{\lambda}}$ is shown in Figure \ref{fig:sparse} for Lactoferrin protein \cite{Norris1991}. The system consists of $10,560$ atoms and is characterized by $10,674$ bonds and $19,249$ harmonic angles. The size of $\mathbf{J}_{\mathbf{\lambda}}$ is therefore $29,923 \times 29,923$. As shown in Figure \ref{fig:sparse}, $\mathbf{J}_{\mathbf{r}_{\alpha}}$ is almost completely sparse even for such a small system.

\subsection{Parallelization} \label{sub:parallel}
The size of the biological systems of interest (such as virus-like particles or assemblies of proteins) makes
MSR a good candidate for parallelization. In the current implementation, MSR
is parallelized for distributed memory systems. This was done with the aid of PETSc
\cite{PETSc-efficient,PETSc-user-ref,PETSc-web-page}, which uses message
passing interface (MPI) \cite{Forum:1994:MMI:898758} to perform linear algebra computations in parallel. The library
supports sparse storage for matrices distributed on multiple nodes. Once the input
coordinates and topology indices are distributed on all processors, the algorithm proceeds by constructing
the RHS of Eqs. (\ref{eq:linear1}-\ref{eq:linear2}) and the Jacobians $\mathbf{J}_{\mathbf{r}_{\alpha}}$,
assembling the Jacobian $\mathbf{J}_{\mathbf{\lambda}}$, and then
solving Eqs. (\ref{eq:linear1}-\ref{eq:linear2}).
A direct Choleski solver was used to solve Eq. (\ref{eq:linear2}) while an iterative solver based on the improved stabilized version of the biconjugate
gradient squared method (IBiCGStab in PETSc) was used to solve Eq. (\ref{eq:linear1}) with the incomplete LU (PCILU in PETSc) chosen as a preconditioner.

\section{Results and Discussion} \label{sec:demo}

\begin{figure}[H]
\centering
\includegraphics[scale=0.5]{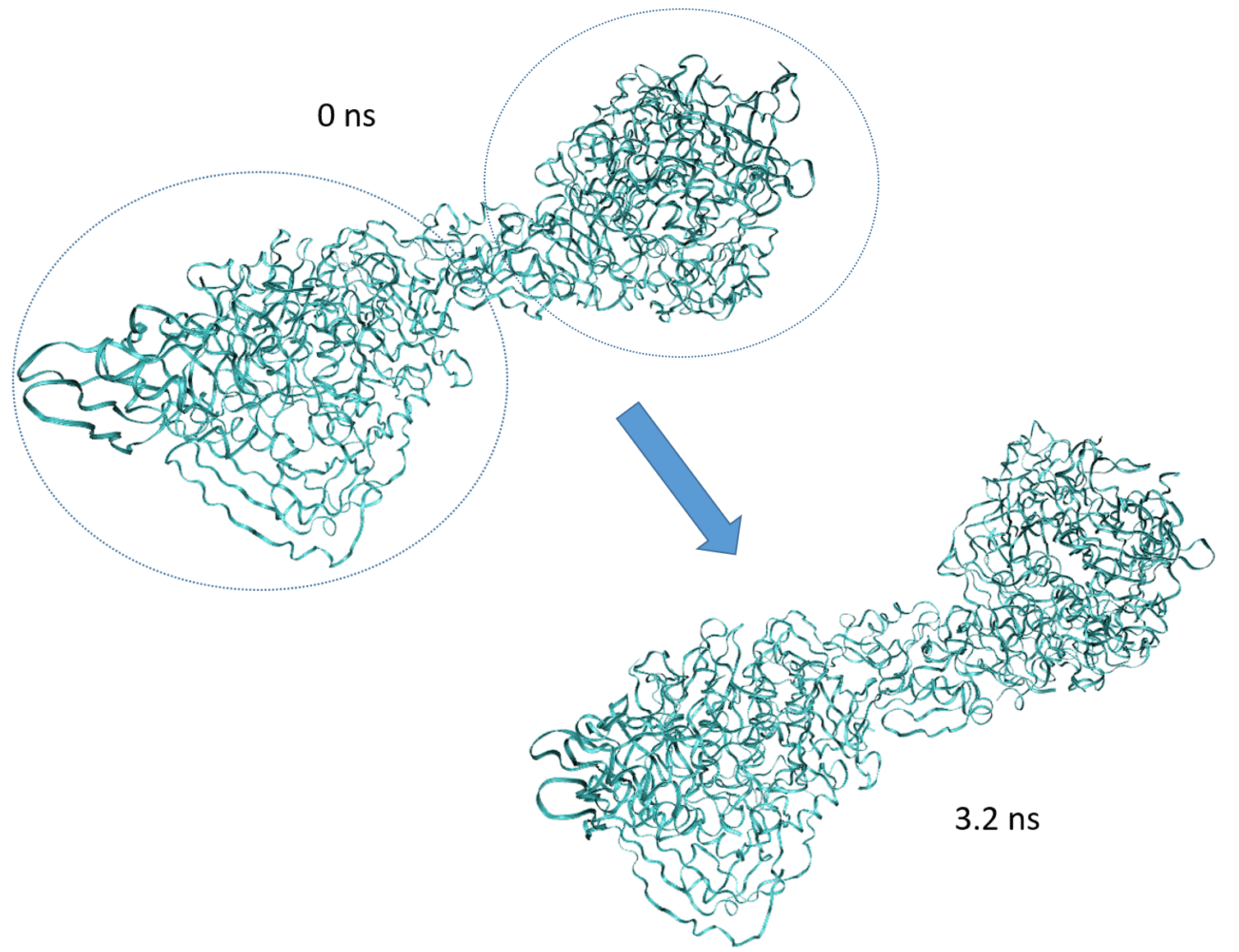}
\caption{Pertussis toxin (PRT) protein undergoes a conformational change under NVT conditions: its two lobes (circled) shrink in time as the protein contracts in aqueous solution of salinity equal to $0.15$ M.}
\label{fig:1PRT}
\end{figure}

 Pertussis toxin (PDB code 1PRT) \cite{1PRT} was used as a demonstration system to assess the accuracy and efficiency of MSR. This protein was simulated under NVT conditions at $300$ K. NaCl counter-ions of concentration $0.15$ M were added for charge neutrality. The system consisted of $603,775$ atoms in a box of dimensions $16$ nm $\times$ $16$ nm $\times$ $24$ nm. An equilibration run with position restraints imposed on the protein was performed for $100$ ps; after thermal equilibrium was established, the system was simulated without any restraints for $3.2$ ns during which the protein underwent a conformational change (Figure \ref{fig:1PRT}).

\subsection{Fine-grained constraint error}
\begin{figure}[H]
\centering
\includegraphics[scale=0.65]{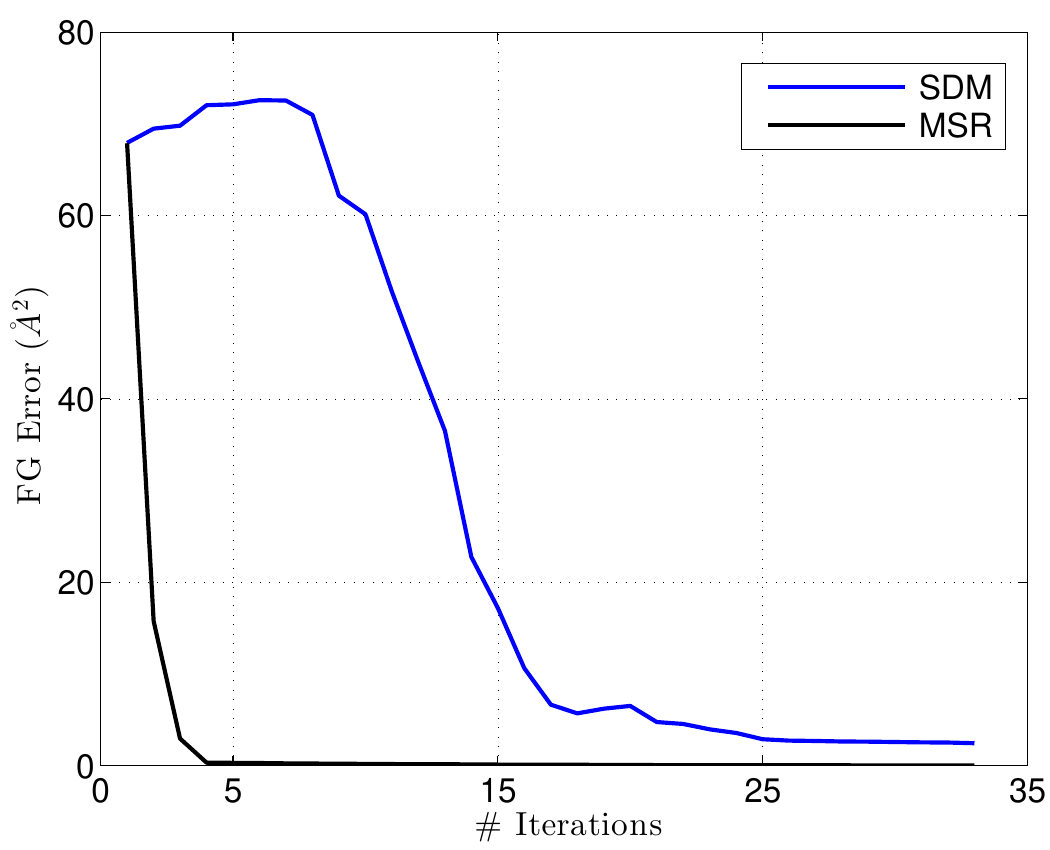}
\caption{Convergence of the fine-grained (FG) error in Eq. (\ref{eq:cons_FS}) is linear. This error drops below $0.1$ \AA\ within $5$ iterations using MSR (in black). In contrast, minimizing the potential energy using the steepest descent method (SDM) takes more iterations to bring the FG error close to $2$ \AA\ (in blue).}
\label{fig:FSerror}
\end{figure}

The error from the FG constraints (Eq. (\ref{eq:cons_FS})) was assessed as follows. First, the space-warping method \cite{Khuloud2002, Joshi2012} was used to coarse-grain the protein at $t=0$ ps using Legendre polynomials of maximum order $k_m=5$. A microstate was then recovered using the method outlined in \cite{Khuloud2002} (Appendix \ref{app:swm}). The recovered microstate was characterized with high bond energies that had to be annealed using extensive energy minimization (the steepest descent method was employed for demonstration), followed by thermalization to bring bond energies to values consistent with the thermal conditions. In contrast, MSR recovers a microstate with modest bond and harmonic angle energies without the need for thermalization and within a fewer number of iterations. This is demonstrated in Figure \ref{fig:FSerror}, which shows the FG error (taken to be the absolute value of the left-hand side of Eq. (\ref{eq:cons_FS})) rapidly vanishes.

\subsection{Coarse-grained constraint error}
\begin{figure}[H]
\centering
\includegraphics[scale=0.65]{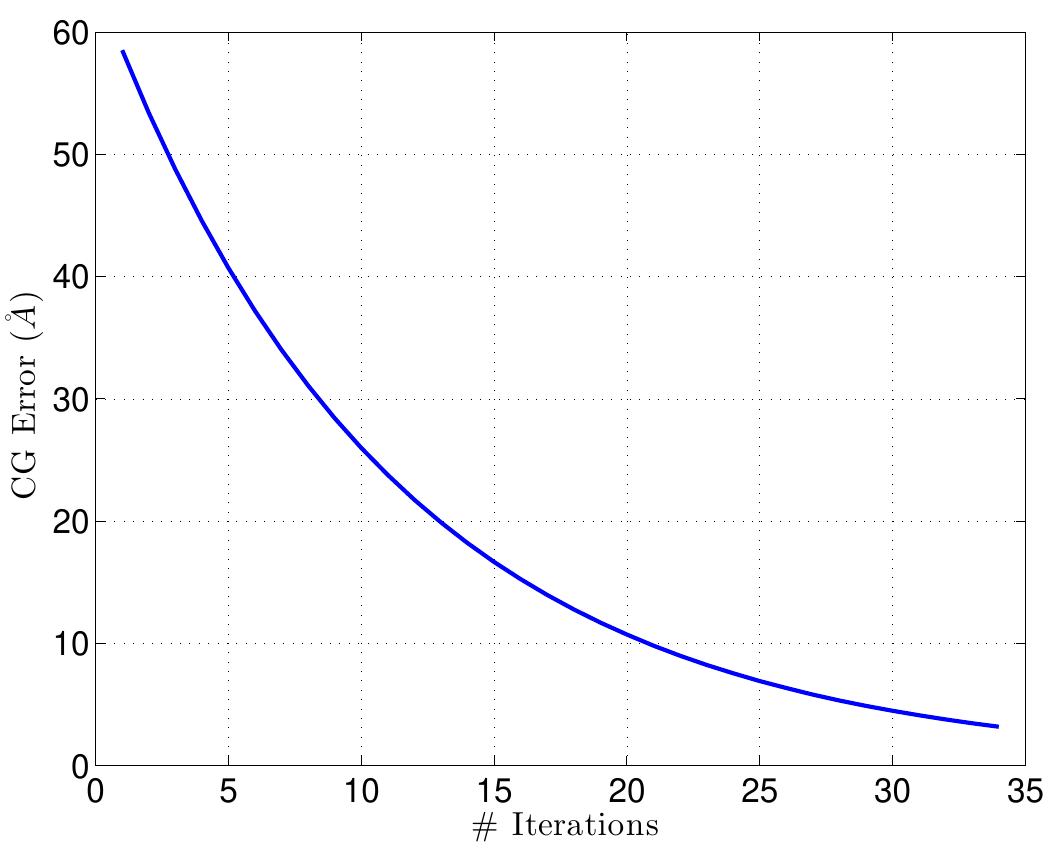}
\caption{In MSR, convergence of the coarse-grained (CG) error in Eq. (\ref{eq:cons_CG}) is linear. This error drops close to $3$ \AA\ in $34$ iterations.}
\label{fig:CGerror}
\end{figure}

\begin{figure}[H]
\centering
\includegraphics[scale=0.65]{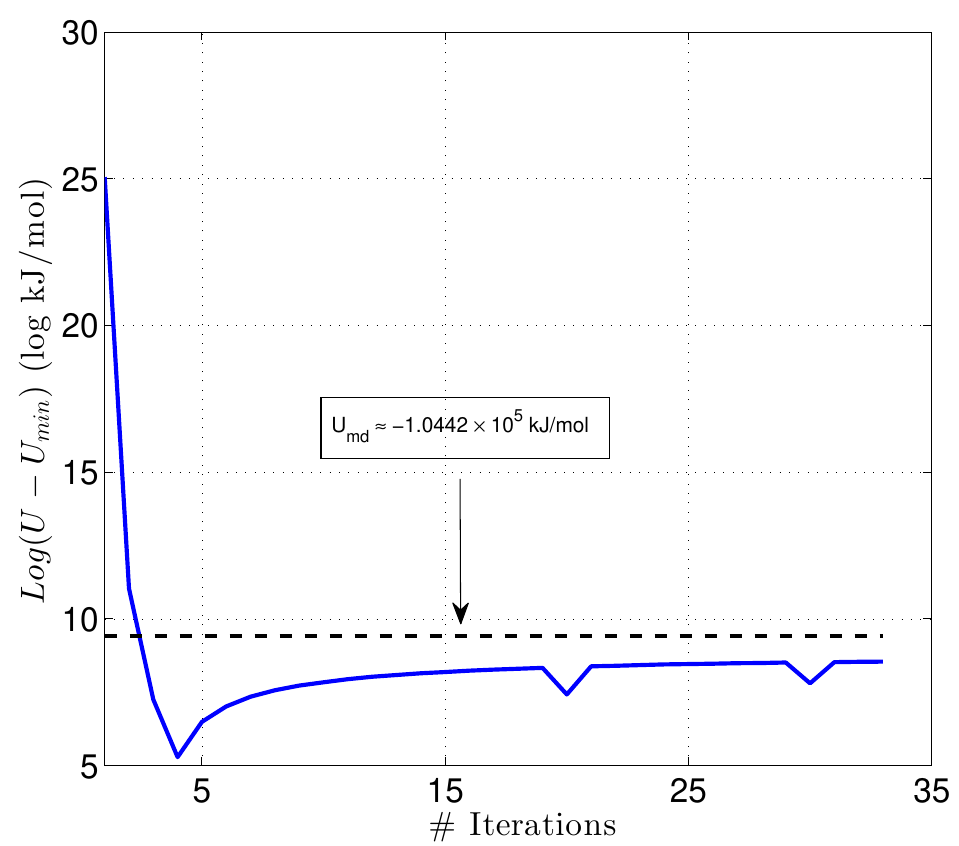}
\caption{The potential energy computed with MSR for pertussis toxin rapidly converges to a value close to that of MD.}
\label{fig:potEnergy}
\end{figure}

Convergence in the error of the CG constraints (Eq. (\ref{eq:cons_CG})) was assessed by taking a microstate of pertussis toxin at $t=10$ ps and introducing noise (a random number between $-1$ and $1$) to all atomic positions. The potential energy of the microstate before it was perturbed was approximately $-1.04 \times 10^5$ kJ/mol, and after perturbation, its potential energy increased to approximately $3.29 \times 10^9$ kJ/mol. The space warping method was then used to coarse-grain the unperturbed microstate using linear Legendre polynomials ($k_m=1$). The constructed CG variables and the bond lengths and harmonic angles computed from MD were then used as input for MSR, which rapidly recovers a microstate consistent with the imposed constraints (Figure \ref{fig:CGerror}). Figure \ref{fig:potEnergy} shows the potential energy difference ($U - U_{min}$) of the recovered microstate is close to that of MD. The reference potantial energy $U_{min}$ was set to $-1.16 \times 10^5$ kJ/mol.

\subsection{Completeness of coarse-grained description}
\begin{figure}[H]
\centering
\includegraphics[scale=0.65]{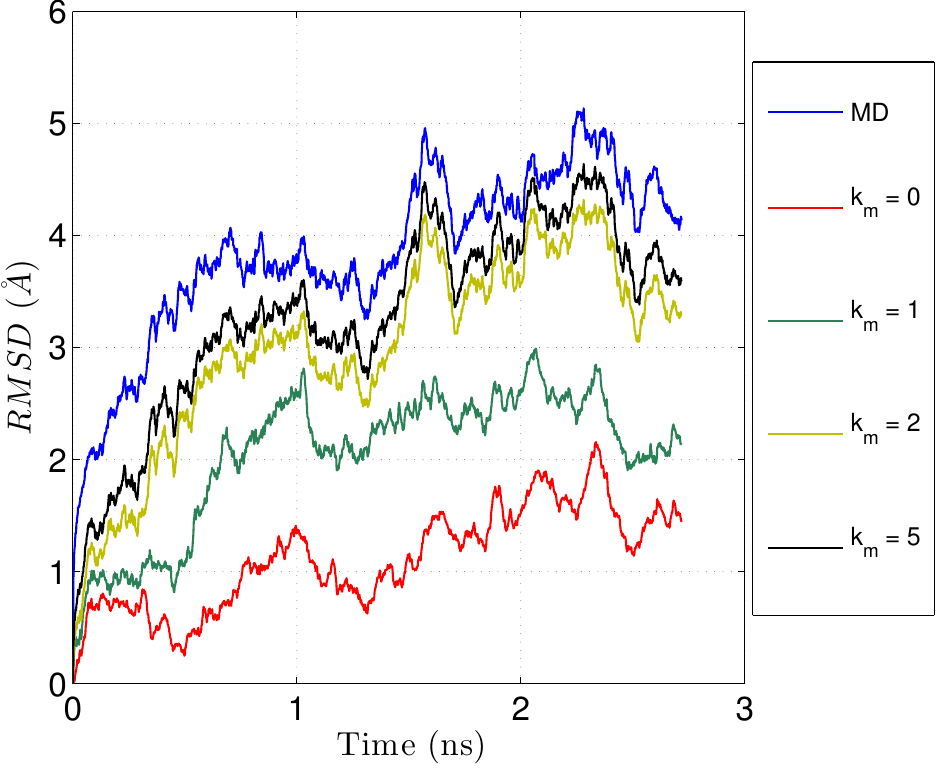}
\caption{The error in RMSD decreases as the number of CG variables increases (represented by $k_m$), indicating that higher order space-warping variables capture finer scale features of the protein.}
\label{fig:kmax}
\end{figure}

\begin{figure}[h]
\centering
\includegraphics[scale=0.6]{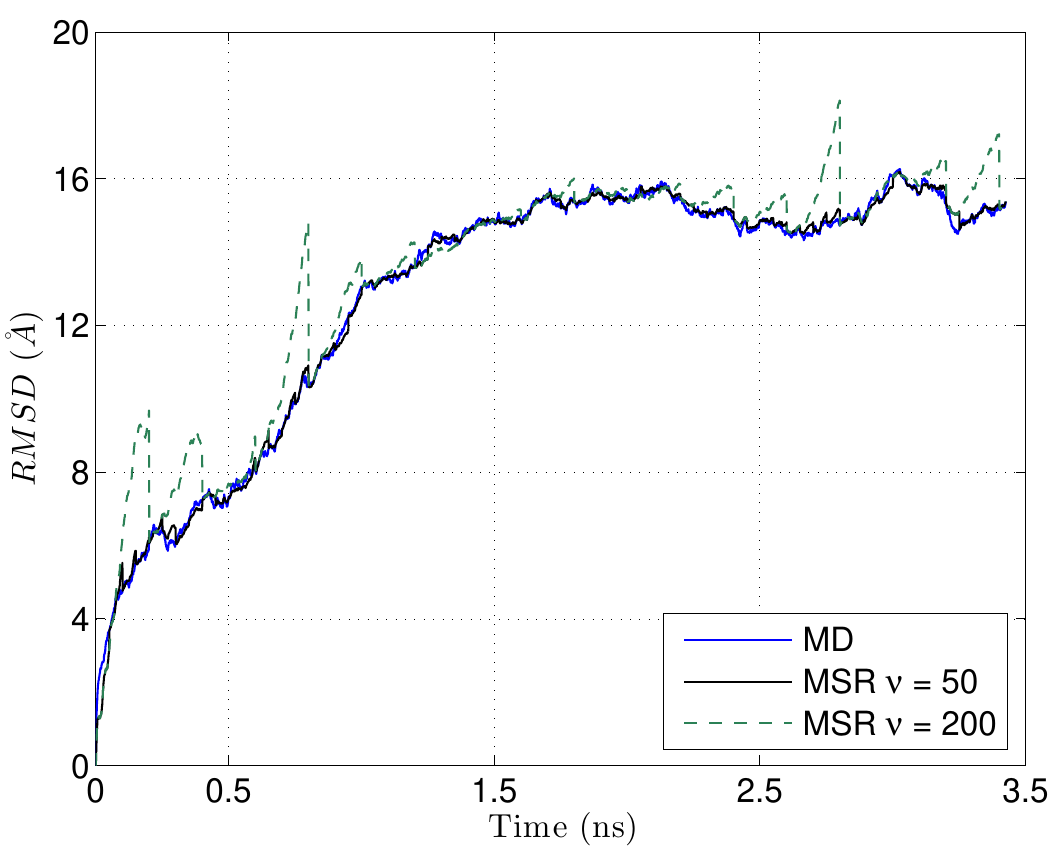}
\caption{MSR reproduces the RMSD of pertussis toxin with relatively small error for $\nu$ = 50. As $\nu$ is increased to $200$, the error in RMSD significantly
increases.}
\label{fig:finegrain}
\end{figure}

In order to analyze the variation in accuracy of MSR due to changes in the number of CG variables included, the RMSD  of pertussis toxin with respect to the initial structure (Figure \ref{fig:kmax}) was computed for various orders and numbers of Legendre polynomials (denoted $k_m$). A time series of the RMSD of the protein was generated using MD, with the structure of each frame aligned to the initial structure. The space-warping variables were then used to coarse-grain the protein, and a new microstate consistent with the CG variables was recovered using MSR and then compared to that obtained from MD. The metric chosen for the fine-graining error is the difference between the RMSD of the protein obtained from MD and that obtained from MSR. As $k_m$ is increased, the number of CG variables increases, and the fine-graining error decreases as expected. However, beyond quadratic Legendre polynomials ($k_m=2$), the rate of convergence in the RMSD error becomes slow for this problem. An alternative way of significantly accelerating this rate is by updating the reference structure. In practice, the latter requires reconstruction of the mapping matrix in Eq. (\ref{eq:dimred}), which can be computationally demanding for large systems represented by a high number of CG variables. Let $\nu$ represent the frequency of updating the reference structure (i.e. the reference structure is updated every $\nu$ CG steps, with each step set to $1$ ps), then the fine-graining error is expected to be proportional to $\nu$. This is demonstrated in Figure \ref{fig:finegrain} which shows the fine-graining error increases as $\nu$ is increased from $50$ to $200$.

\subsection{Scalability}
\begin{figure}[H]
\centering
\includegraphics[scale=0.5]{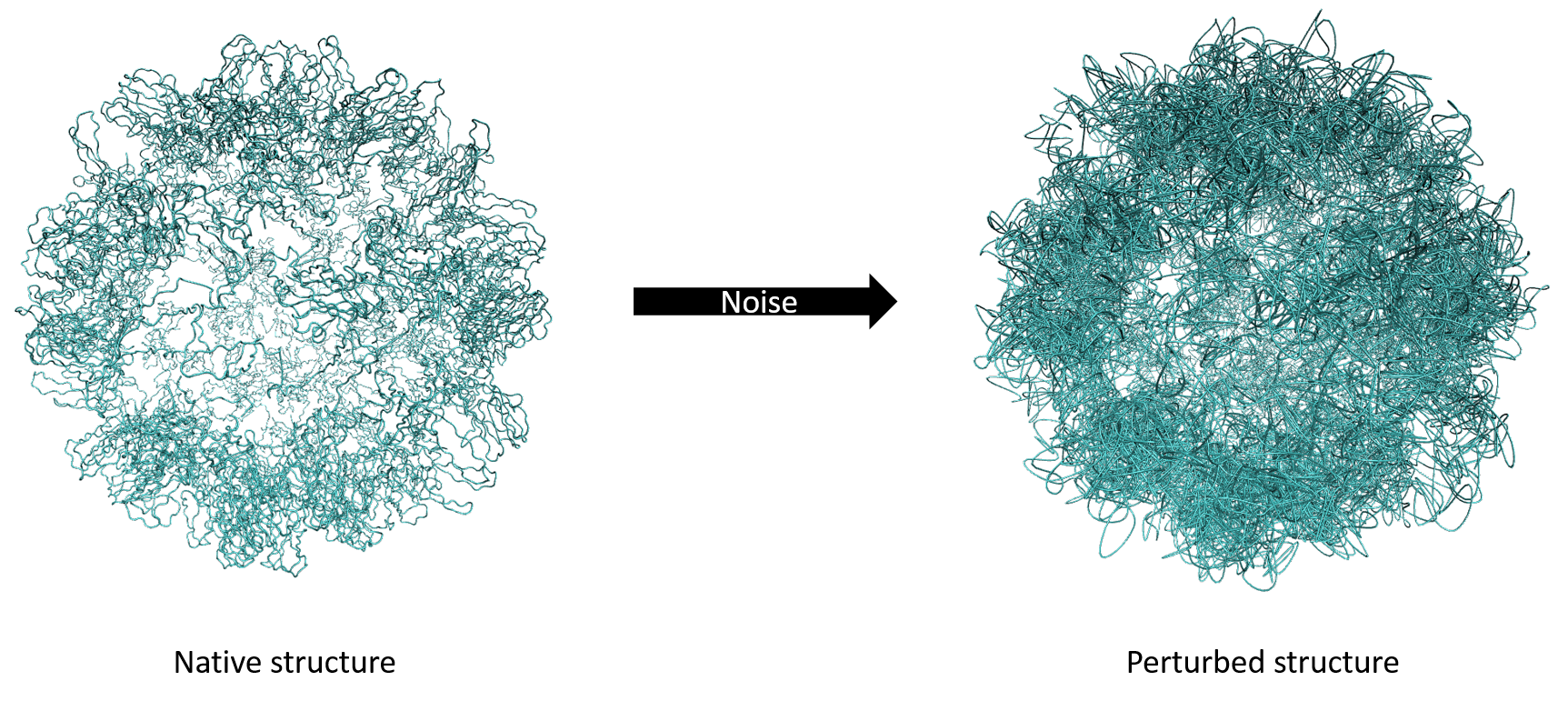}
\caption{A snapshot of the Cowpea Chlorotic Mottle virus capsid in its native (left) and perturbed (right) forms.}
\label{fig:CCMV}
\end{figure}

\begin{figure}[H]
\centering
	\includegraphics[scale=0.6]{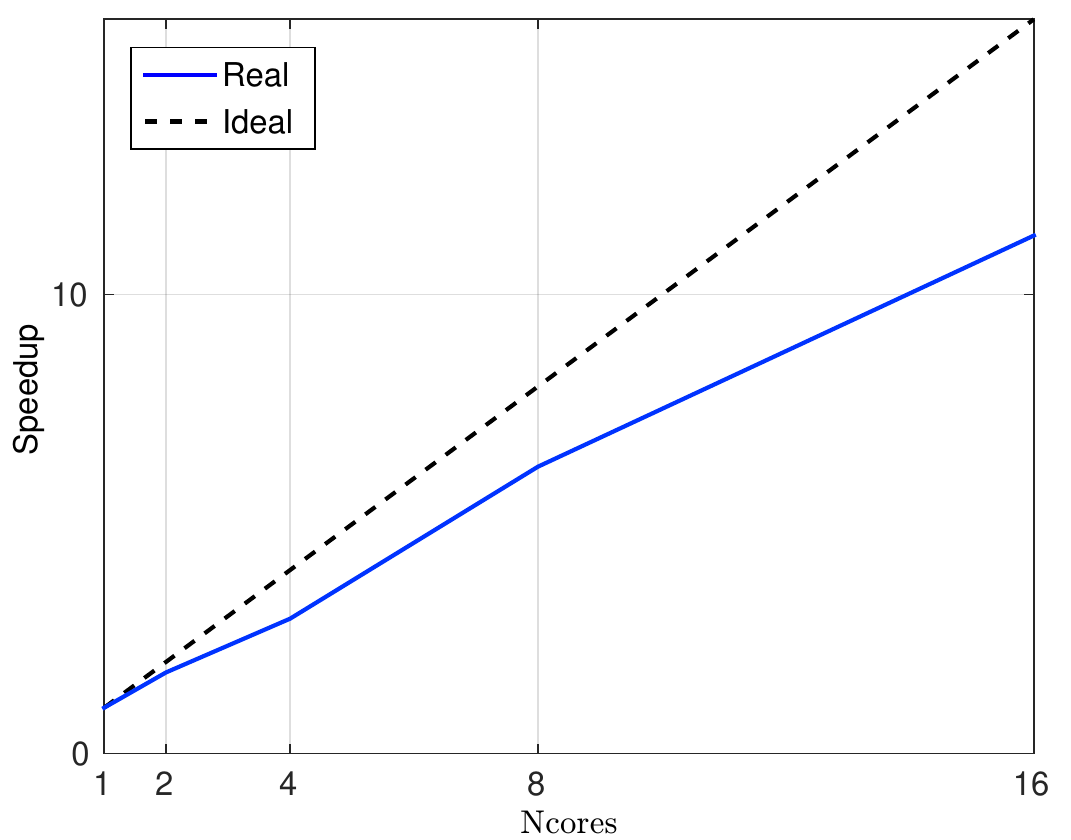}
	\caption{Strong scalability of MSR is shown for the Human Papillomavirus capsid using a total of $16$ cores. The measured speedup is with respect to one core (serial run) based on the CPU time.}
	\label{fig:scalability}
\end{figure}

To analyze the parallel performance and scalability of MSR, a Cowpea Chlorotic Mottle virus capsid (PDB code 1CWP) was used as a demonstration system (Fig. (\ref{fig:CCMV})). The capsid was generated using BIOMT transformations \cite{CCMV}. This virus-like particle (VLP) supports a structure consisting of $450,840$ atoms. Linear space warping variables (using $k_m=1$) were then computed for the VLP before it was perturbed by adding noise (random number between $-10$ and $10$) to all of its atomic positions. The microstate was then recovered in $10$ MSR iterations for a total of $1,348,680$ FG constraints using an increasing number of cores. Simulations were performed on Indiana University's Karst cluster, using a dual Intel Xeon E5-2650 v2 $8$-core processor and a total of $32$ GB of RAM. MSR shows promising strong scalability over a total of $16$ cores (Fig. (\ref{fig:scalability})). Further optimization of the algorithm using OpenMP or GPU-based acceleration should lead to higher speedups.

\section{Conclusion} \label{sec:conclusion}
Microstate reconstruction is a key element of multiscale MD algorithms. MSR, an efficient method that constructs microstates consistent with the evolved CG description and thermal conditions is presented and demonstrated for proteins and their assemblies. Using these microstates to generate dynamical information needed to update the CG state in MF yields accurate and efficient multiscale simulation of molecular systems. MSR can be further improved by taking VdW interactions into account when reconstructing the microstate. This can be achieved by incoporating historical information (such as several microstates obtained from MD) into the optimization problem. The numerical implementation of MSR leads to highly sparse matrices; consequently, the prallel implementation shows good scaling with the size of the simulated systems. This suggests that the algorithm is suitable for supramillion-atom systems such as virus-like particles and other nanomaterials.

\begin{acknowledgement}
This work was supported by in part by the NSF division of material science research (grant 1533988, $80\%$) and the Indiana Clinical and Translational Sciences Institute, funded in part by grant UL1 TR001108 from the National Institutes of Health, National Center for Advancing Translational Sciences, Clinical and Translational Sciences Award ($20\%$). The authors also acknowledge support by the IU College of Arts and Sciences via the Center for Theoretical and Computational Nanoscience, and Lilly Endowment, Inc., through its support for the Indiana University Pervasive Technology Institute, and in part by the Indiana METACyt Initiative. The Indiana METACyt Initiative at IU is also supported in part by Lilly Endowment, Inc. The content is solely the responsibility of the authors and does not necessarily represent the official views of the National Institutes of Health.
\end{acknowledgement}

\appendix
\section{The space-warping method} \label{app:swm}
\begin{figure}[H]
\centering
\includegraphics[scale=0.2]{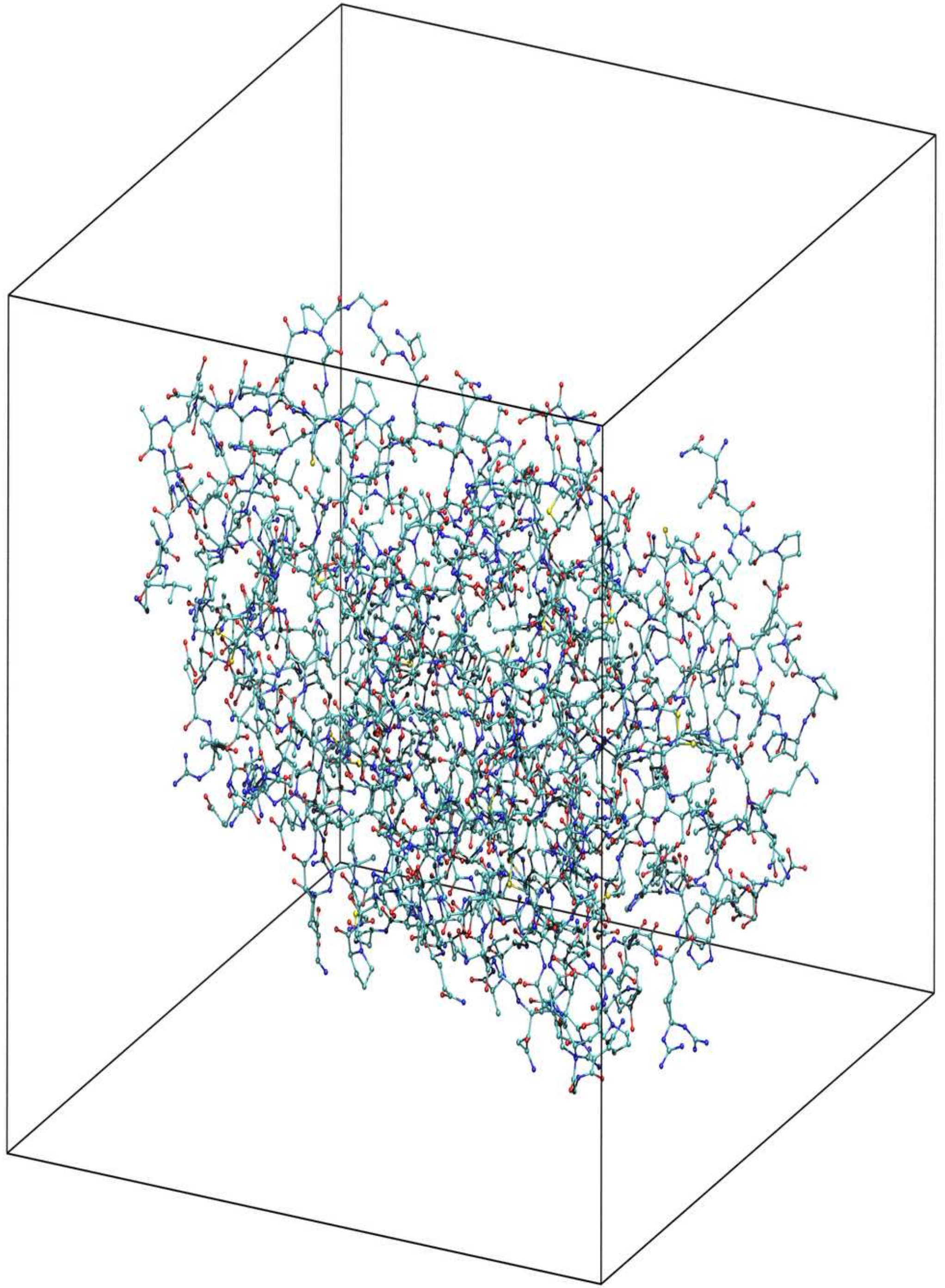}
\caption{The space warping method embeds a macromolecule in a normalized orthogonal box in which the basis functions ($\mathbf{B}$) are constructed. }
\label{fig:app-box}
\end{figure}

The space-warping method is a coarse-graining method suitable
for representing macromolecules in low-dimensional manifolds \cite{Khuloud2002,Joshi2012}. The development starts by writing the $N-$atom coordinates (denoted $\mathbf{r}$) in terms of a set of coarse-grained variables (denoted $\mathbf{\phi}$) via a
Fourier-like expansion:
\begin{equation}
\mathbf{r}_i = \mathbf{r}_c^0 + \sum_{\underline{k}} \mathbf{B}_{\underline{k}}(\mathbf{r}_{i}^{0}) \mathbf{\phi}_{\underline{k}} + \mathbf{\sigma}_{i},
\label{eq:micro}
\end{equation}
where $\underline{k}$ is a triplet of indices ($k_x$, $k_y$, and $k_z$) that vary between $0$ and $N_{CG}$, and the `Fourier modes' of order $\underline{k}$ are represented by $\mathbf{\phi}_{\underline{k}}$, a $3-$dimensional vector which serves as a CG variable that captures large-scale macromolecular conformational changes;
$\mathbf{r}_i$ is the position of atom $i$; $\mathbf{B}$ is a matrix (of size $N \times N_{CG}$) of the product of three Legendre functions of orders
$k_x$, $k_y$, and $k_z$ along the $x$, $y$, and $z$ axes, respectively; this matrix depends on the positions of a reference all-atom configuration
(denoted $\mathbf{r}^{0}$) of center of mass designated $\mathbf{r}_c^0$; $\mathbf{\sigma}_i$ is a $3-$dimensional vector of residual displacements that result from the difference between the positions generated by the coherent deformations. The Legendre polynomials are constructed over a normalized orthogonal box shown in Figure \ref{fig:app-box}.

Coarse-graining is achieved by mass-weighted least square minimization, i.e. by minimizing $\sum_{i=1}^N m_i \mathbf{\sigma}_i^2$ with respect
to $\mathbf{\phi}_{\underline{k}}$, with $m_i$ being the mass of atom $i$. The result is a set of CG variables that serve as generalized centers of mass:
\begin{equation}
\mathbf{B}^T \mathbf{M} \mathbf{B} \mathbf{\phi}_{\alpha} =  \mathbf{B}^T \mathbf{M} \mathbf{r}_{\alpha},
\end{equation}
where $\mathbf{M}$ is a diagonal matrix of the atomic masses, $\mathbf{\phi}_{\alpha}$ is a vector of CG coordinates of order $\underline{k}$ for the $\alpha$ axis, and $\mathbf{r}_{\alpha}$ is a vector of atomic positions (minus the center of mass of the macromolecule) for the $\alpha$ axis. Thus, the coarse-graining matrix $\mathbf{Q}$ introduced in Eq. (\ref{eq:dimred}) is
$ \left(\mathbf{B}^T \mathbf{M} \mathbf{B} \right)^{-1} \mathbf{B}^T \mathbf{M}$.  

The total order of the method is designated $k_m$ such that $k_m \geq k_x + k_y +k_z$. For example, if $k_m = 0$, then $\{k_x,k_y,k_z\} = \{0,0,0\}$. The total number of CG variables in this case is $3 \times 1$, corresponding to $\mathbf{\phi}_{000}$ being the center of mass of the macromolecule. As $k_m$ increases, the CG variables capture additional information from the atomic scale, but they vary less slowly in time. Therefore, the space warping CG variables are classified into low order and high order variables. The former characterize the larger scale disturbances, while the latter capture short-scale ones \cite{Singharoy2011,Joshi2012}. For $k_m=1$, $\{k_x,k_y,k_z\} = \{0,0,0\}, \{1,0,0\},\{0,1,0\},\{0,0,1\}$. In this case, the total number of CG variables is $3 \times 4$. It can be shown for $k_m=1$, $\sum_{i=1}^{N} \mathbf{Q}_{k_xk_yk_z}(\mathbf{r}_{i}^{0})$ is equal to $1$ if $k_x = k_y = k_z = 0$ and $0$ otherwise. Thus, if the atomistic system has translated a distance $d$ in all three directions, then this translation is captured by $\phi_{000}$ since $\Delta \phi = \mathbf{Q} \Delta \mathbf{r} = d \mathbf{Q} \mathbf{1}$, the first component of which is $d$, while the rest are $0$. It can be further shown that $\phi_{100}$, $\phi_{010}$, and $\phi_{001}$ capture rotational motion \cite{Khuloud2002}, while 2nd order CG variables ($k_x + k_y + k_z =2$) capture non-linear transformations such as bending that macromolecular systems undergo. 

\section{Regularization of inverse problems} \label{app:regular}
Inverse problems are often ill-posed \cite{hansen1998rank}. In the context of reverse coarse-graining, solving such problems can be numerically challenging because the solution might exhibit numerical instabilities. For instance, if least square minimization is employed for recovering an all-atom state from the CG description using Eq. (\ref{eq:dimred}), the reverse map $\mathbf{Q}^T\mathbf{Q}$ amplifies the high-frequency noise, thus leading to numerical instabilities. Therefore, in practice regularization is performed by including additional constraints in the optimization problem. The most commonly used regularization is the $L_2$ norm of the solution vector (or in the present context, the vector of atomic positions $\mathbf{r}_{\alpha}$). In MSR, regularization is cast in terms of: 1) the difference between the atomic positions, $\mathbf{r}_{\alpha}$, and the reference atomic positions, $\mathbf{r}^0_{\alpha}$, and 2) the Lagrange multipliers that enforce specific constraints on the FG constraints in Eq. (\ref{eq:optimization2}).

\bibliography{AndrewBib}

\providecommand{\latin}[1]{#1}
\providecommand*\mcitethebibliography{\thebibliography}
\csname @ifundefined\endcsname{endmcitethebibliography}
  {\let\endmcitethebibliography\endthebibliography}{}
\begin{mcitethebibliography}{38}
\providecommand*\natexlab[1]{#1}
\providecommand*\mciteSetBstSublistMode[1]{}
\providecommand*\mciteSetBstMaxWidthForm[2]{}
\providecommand*\mciteBstWouldAddEndPuncttrue
  {\def\EndOfBibitem{\unskip.}}
\providecommand*\mciteBstWouldAddEndPunctfalse
  {\let\EndOfBibitem\relax}
\providecommand*\mciteSetBstMidEndSepPunct[3]{}
\providecommand*\mciteSetBstSublistLabelBeginEnd[3]{}
\providecommand*\EndOfBibitem{}
\mciteSetBstSublistMode{f}
\mciteSetBstMaxWidthForm{subitem}{(\alph{mcitesubitemcount})}
\mciteSetBstSublistLabelBeginEnd
  {\mcitemaxwidthsubitemform\space}
  {\relax}
  {\relax}

\bibitem[Ortoleva(2005)]{Ortoleva2005}
Ortoleva,~P.~J. \emph{J. Chem. Phys.} \textbf{2005}, \emph{109},
  2770--2771\relax
\mciteBstWouldAddEndPuncttrue
\mciteSetBstMidEndSepPunct{\mcitedefaultmidpunct}
{\mcitedefaultendpunct}{\mcitedefaultseppunct}\relax
\EndOfBibitem
\bibitem[Pankavich \latin{et~al.}(2008)Pankavich, Shreif, and
  Ortoleva]{Ortoleva2008}
Pankavich,~S.; Shreif,~Z.; Ortoleva,~P.~J. \emph{Physica A} \textbf{2008},
  \emph{387}, 4053--4069\relax
\mciteBstWouldAddEndPuncttrue
\mciteSetBstMidEndSepPunct{\mcitedefaultmidpunct}
{\mcitedefaultendpunct}{\mcitedefaultseppunct}\relax
\EndOfBibitem
\bibitem[Pankavich \latin{et~al.}(2009)Pankavich, Shreif, Miao, and
  Ortoleva]{Ortoleva2009}
Pankavich,~S.; Shreif,~Z.; Miao,~Y.; Ortoleva,~P.~J. \emph{J. Chem. Phys.}
  \textbf{2009}, \emph{130}, 194115--194124\relax
\mciteBstWouldAddEndPuncttrue
\mciteSetBstMidEndSepPunct{\mcitedefaultmidpunct}
{\mcitedefaultendpunct}{\mcitedefaultseppunct}\relax
\EndOfBibitem
\bibitem[Cheluvaraja and Ortoleva(2010)Cheluvaraja, and
  Ortoleva]{Cheluvaraja2010}
Cheluvaraja,~S.; Ortoleva,~P.~J. \emph{J. Chem. Phys.} \textbf{2010},
  \emph{132}, 75102--75110\relax
\mciteBstWouldAddEndPuncttrue
\mciteSetBstMidEndSepPunct{\mcitedefaultmidpunct}
{\mcitedefaultendpunct}{\mcitedefaultseppunct}\relax
\EndOfBibitem
\bibitem[Abi~Mansour and Ortoleva(2016)Abi~Mansour, and
  Ortoleva]{abi2016implicit}
Abi~Mansour,~A.; Ortoleva,~P.~J. \emph{Journal of chemical theory and
  computation} \textbf{2016}, \relax
\mciteBstWouldAddEndPunctfalse
\mciteSetBstMidEndSepPunct{\mcitedefaultmidpunct}
{}{\mcitedefaultseppunct}\relax
\EndOfBibitem
\bibitem[Theodoropoulos \latin{et~al.}(2000)Theodoropoulos, Qian, and
  Kevrekidis]{Kevrekidis2000}
Theodoropoulos,~C.; Qian,~Y.-H.; Kevrekidis,~I.~G. \emph{Proc. Natl. Acad.
  Sci.} \textbf{2000}, \emph{97}, 9840--9843\relax
\mciteBstWouldAddEndPuncttrue
\mciteSetBstMidEndSepPunct{\mcitedefaultmidpunct}
{\mcitedefaultendpunct}{\mcitedefaultseppunct}\relax
\EndOfBibitem
\bibitem[Gear \latin{et~al.}(2002)Gear, Kevrekidis, and
  Theodoropoulos]{Gear2002}
Gear,~C.~W.; Kevrekidis,~I.~G.; Theodoropoulos,~C. \emph{Comput. Chem. Eng.}
  \textbf{2002}, \emph{26}, 941--963\relax
\mciteBstWouldAddEndPuncttrue
\mciteSetBstMidEndSepPunct{\mcitedefaultmidpunct}
{\mcitedefaultendpunct}{\mcitedefaultseppunct}\relax
\EndOfBibitem
\bibitem[Gear \latin{et~al.}(2003)Gear, Hyman, Kevrekidid, Kevrekidis, Runborg,
  and Theodoropoulos]{Gear2003}
Gear,~C.~W.; Hyman,~J.~M.; Kevrekidid,~P.~G.; Kevrekidis,~I.~G.; Runborg,~O.;
  Theodoropoulos,~C. \emph{Commun. Math. Sci.} \textbf{2003}, \emph{1},
  715--762\relax
\mciteBstWouldAddEndPuncttrue
\mciteSetBstMidEndSepPunct{\mcitedefaultmidpunct}
{\mcitedefaultendpunct}{\mcitedefaultseppunct}\relax
\EndOfBibitem
\bibitem[Kevrekidis and Samaey(2009)Kevrekidis, and Samaey]{Kevrekidis2009}
Kevrekidis,~I.; Samaey,~G. \emph{Annu. Rev. Phys. Chem.} \textbf{2009},
  \emph{60}, 321--344\relax
\mciteBstWouldAddEndPuncttrue
\mciteSetBstMidEndSepPunct{\mcitedefaultmidpunct}
{\mcitedefaultendpunct}{\mcitedefaultseppunct}\relax
\EndOfBibitem
\bibitem[Bahar \latin{et~al.}(1997)Bahar, Atilgan, and Erman]{Bahar1997}
Bahar,~I.; Atilgan,~R.~A.; Erman,~B. \emph{Folding and Design} \textbf{1997},
  \emph{2}, 173--181\relax
\mciteBstWouldAddEndPuncttrue
\mciteSetBstMidEndSepPunct{\mcitedefaultmidpunct}
{\mcitedefaultendpunct}{\mcitedefaultseppunct}\relax
\EndOfBibitem
\bibitem[Noid \latin{et~al.}(2008)Noid, Chu, Ayton, Krishna, Izvekov, Voth,
  Das, and Andersen]{Voth2008}
Noid,~W.~G.; Chu,~J.-W.; Ayton,~G.~S.; Krishna,~V.; Izvekov,~S.; Voth,~G.~A.;
  Das,~A.; Andersen,~H.~C. \emph{J. Chem. Phys.} \textbf{2008}, \emph{128},
  244114--244124\relax
\mciteBstWouldAddEndPuncttrue
\mciteSetBstMidEndSepPunct{\mcitedefaultmidpunct}
{\mcitedefaultendpunct}{\mcitedefaultseppunct}\relax
\EndOfBibitem
\bibitem[Reith \latin{et~al.}(2003)Reith, Putz, and Muller-Plathe]{Reith2003}
Reith,~D.; Putz,~M.; Muller-Plathe,~F. \emph{J. Comput. Chem.} \textbf{2003},
  \emph{24}, 1624--1636\relax
\mciteBstWouldAddEndPuncttrue
\mciteSetBstMidEndSepPunct{\mcitedefaultmidpunct}
{\mcitedefaultendpunct}{\mcitedefaultseppunct}\relax
\EndOfBibitem
\bibitem[Shiha \latin{et~al.}(2006)Shiha, Arkhipov, Freddolino, and
  Schulten]{Schulten2006}
Shiha,~A.~Y.; Arkhipov,~A.; Freddolino,~P.~L.; Schulten,~K. \emph{J. Phys.
  Chem. B} \textbf{2006}, \emph{110}, 3674--3684\relax
\mciteBstWouldAddEndPuncttrue
\mciteSetBstMidEndSepPunct{\mcitedefaultmidpunct}
{\mcitedefaultendpunct}{\mcitedefaultseppunct}\relax
\EndOfBibitem
\bibitem[Muller-Plathe(2002)]{Muller2002}
Muller-Plathe,~F. \emph{ChemPhysChem.} \textbf{2002}, \emph{3}, 754--769\relax
\mciteBstWouldAddEndPuncttrue
\mciteSetBstMidEndSepPunct{\mcitedefaultmidpunct}
{\mcitedefaultendpunct}{\mcitedefaultseppunct}\relax
\EndOfBibitem
\bibitem[Rudd and Broughton(1998)Rudd, and Broughton]{Broughton1998}
Rudd,~R.~E.; Broughton,~J.~Q. \emph{Phys. Rev. B} \textbf{1998}, \emph{58},
  5893--5896\relax
\mciteBstWouldAddEndPuncttrue
\mciteSetBstMidEndSepPunct{\mcitedefaultmidpunct}
{\mcitedefaultendpunct}{\mcitedefaultseppunct}\relax
\EndOfBibitem
\bibitem[Mansour and Ortoleva(2014)Mansour, and Ortoleva]{AbiMansour2014}
Mansour,~A.~A.; Ortoleva,~P. \emph{J. Chem. Theory Comput.} \textbf{2014},
  \emph{10}, 518--523\relax
\mciteBstWouldAddEndPuncttrue
\mciteSetBstMidEndSepPunct{\mcitedefaultmidpunct}
{\mcitedefaultendpunct}{\mcitedefaultseppunct}\relax
\EndOfBibitem
\bibitem[Sereda \latin{et~al.}(2014)Sereda, Espinosa-Duran, and
  Ortoleva]{Sereda2014}
Sereda,~Y.~V.; Espinosa-Duran,~J.~M.; Ortoleva,~P.~J. \emph{J. Chem. Phys.}
  \textbf{2014}, \emph{140}, 074102\relax
\mciteBstWouldAddEndPuncttrue
\mciteSetBstMidEndSepPunct{\mcitedefaultmidpunct}
{\mcitedefaultendpunct}{\mcitedefaultseppunct}\relax
\EndOfBibitem
\bibitem[Singharoy \latin{et~al.}(2011)Singharoy, Cheluvaraja, and
  Ortoleva]{Singharoy2011}
Singharoy,~A.; Cheluvaraja,~S.; Ortoleva,~P.~J. \emph{J. Chem. Phys.}
  \textbf{2011}, \emph{134}, 44104--44120\relax
\mciteBstWouldAddEndPuncttrue
\mciteSetBstMidEndSepPunct{\mcitedefaultmidpunct}
{\mcitedefaultendpunct}{\mcitedefaultseppunct}\relax
\EndOfBibitem
\bibitem[Singharoy \latin{et~al.}(2012)Singharoy, Sereda, and
  Ortoleva]{Singharoy2012}
Singharoy,~A.; Sereda,~Y.; Ortoleva,~P.~J. \emph{J. Chem. Theory Comput.}
  \textbf{2012}, \emph{8}, 1379--1392\relax
\mciteBstWouldAddEndPuncttrue
\mciteSetBstMidEndSepPunct{\mcitedefaultmidpunct}
{\mcitedefaultendpunct}{\mcitedefaultseppunct}\relax
\EndOfBibitem
\bibitem[Hansen(1998)]{hansen1998rank}
Hansen,~P.~C. \emph{Rank-deficient and Discrete Ill-posed Problems: Numerical
  Aspects of Linear Inversion}; Siam, 1998; Vol.~4\relax
\mciteBstWouldAddEndPuncttrue
\mciteSetBstMidEndSepPunct{\mcitedefaultmidpunct}
{\mcitedefaultendpunct}{\mcitedefaultseppunct}\relax
\EndOfBibitem
\bibitem[Ensing and Nielsen(2010)Ensing, and Nielsen]{Ensing2010}
Ensing,~B.; Nielsen,~S.~O. In \emph{Trends in Computational Nanomechanics:
  Transcending Length and Time Scales}; Dumitrica,~T., Ed.; Springer
  Netherlands: Dordrecht, 2010; Chapter Multiscale Molecular Dynamics and the
  Reverse Mapping Problem, pp 25--59\relax
\mciteBstWouldAddEndPuncttrue
\mciteSetBstMidEndSepPunct{\mcitedefaultmidpunct}
{\mcitedefaultendpunct}{\mcitedefaultseppunct}\relax
\EndOfBibitem
\bibitem[Somogyi \latin{et~al.}(2016)Somogyi, Mansour, and Ortoleva]{protoMD}
Somogyi,~E.; Mansour,~A.~A.; Ortoleva,~P.~J. \emph{Comput. Phys. Commun.}
  \textbf{2016}, \emph{202}, 337 -- 350\relax
\mciteBstWouldAddEndPuncttrue
\mciteSetBstMidEndSepPunct{\mcitedefaultmidpunct}
{\mcitedefaultendpunct}{\mcitedefaultseppunct}\relax
\EndOfBibitem
\bibitem[Balay \latin{et~al.}(1997)Balay, Gropp, McInnes, and
  Smith]{PETSc-efficient}
Balay,~S.; Gropp,~W.~D.; McInnes,~L.~C.; Smith,~B.~F. Efficient Management of
  Parallelism in Object Oriented Numerical Software Libraries. Modern Software
  Tools in Scientific Computing. 1997; pp 163--202\relax
\mciteBstWouldAddEndPuncttrue
\mciteSetBstMidEndSepPunct{\mcitedefaultmidpunct}
{\mcitedefaultendpunct}{\mcitedefaultseppunct}\relax
\EndOfBibitem
\bibitem[Balay \latin{et~al.}(2013)Balay, Brown, Buschelman, Eijkhout, Gropp,
  Kaushik, Knepley, McInnes, Smith, and Zhang]{PETSc-user-ref}
Balay,~S.; Brown,~J.; Buschelman,~K.; Eijkhout,~V.; Gropp,~W.~D.; Kaushik,~D.;
  Knepley,~M.~G.; McInnes,~L.~C.; Smith,~B.~F.; Zhang,~H. \emph{{PETS}c Users
  Manual}; 2013\relax
\mciteBstWouldAddEndPuncttrue
\mciteSetBstMidEndSepPunct{\mcitedefaultmidpunct}
{\mcitedefaultendpunct}{\mcitedefaultseppunct}\relax
\EndOfBibitem
\bibitem[Balay \latin{et~al.}(2013)Balay, Brown, Buschelman, Gropp, Kaushik,
  Knepley, McInnes, Smith, and Zhang]{PETSc-web-page}
Balay,~S.; Brown,~J.; Buschelman,~K.; Gropp,~W.~D.; Kaushik,~D.;
  Knepley,~M.~G.; McInnes,~L.~C.; Smith,~B.~F.; Zhang,~H. PETSc Web page. 2013;
  http://www.mcs.anl.gov/petsc\relax
\mciteBstWouldAddEndPuncttrue
\mciteSetBstMidEndSepPunct{\mcitedefaultmidpunct}
{\mcitedefaultendpunct}{\mcitedefaultseppunct}\relax
\EndOfBibitem
\bibitem[Beazley(1996)]{SWIG}
Beazley,~D.~M. SWIG: An Easy to Use Tool for Integrating Scripting Languages
  with C and C++. Proceedings of the 4th Conference on USENIX Tcl/Tk Workshop,
  1996 - Volume 4. Berkeley, CA, USA, 1996; pp 15--15\relax
\mciteBstWouldAddEndPuncttrue
\mciteSetBstMidEndSepPunct{\mcitedefaultmidpunct}
{\mcitedefaultendpunct}{\mcitedefaultseppunct}\relax
\EndOfBibitem
\bibitem[Abi-Mansour()]{MSR}
Abi-Mansour,~A. MSR. {@ONLINE}. \url{https://github.com/CTCNano/MSR}\relax
\mciteBstWouldAddEndPuncttrue
\mciteSetBstMidEndSepPunct{\mcitedefaultmidpunct}
{\mcitedefaultendpunct}{\mcitedefaultseppunct}\relax
\EndOfBibitem
\bibitem[Jaqaman and Ortoleva(2002)Jaqaman, and Ortoleva]{Khuloud2002}
Jaqaman,~K.; Ortoleva,~P.~J. \emph{J. Comput. Chem.} \textbf{2002}, \emph{23},
  484--491\relax
\mciteBstWouldAddEndPuncttrue
\mciteSetBstMidEndSepPunct{\mcitedefaultmidpunct}
{\mcitedefaultendpunct}{\mcitedefaultseppunct}\relax
\EndOfBibitem
\bibitem[Singharoy \latin{et~al.}(2012)Singharoy, Joshi, Miao, and
  Ortoleva]{Joshi2012}
Singharoy,~A.; Joshi,~H.; Miao,~Y.; Ortoleva,~P.~J. \emph{J. Phys. Chem. B}
  \textbf{2012}, \emph{116}, 8423--8434\relax
\mciteBstWouldAddEndPuncttrue
\mciteSetBstMidEndSepPunct{\mcitedefaultmidpunct}
{\mcitedefaultendpunct}{\mcitedefaultseppunct}\relax
\EndOfBibitem
\bibitem[Nocedal and Wright(2006)Nocedal, and Wright]{Optimization2006}
Nocedal,~J.; Wright,~S. \emph{Numerical Optimization}; Springer, 2006\relax
\mciteBstWouldAddEndPuncttrue
\mciteSetBstMidEndSepPunct{\mcitedefaultmidpunct}
{\mcitedefaultendpunct}{\mcitedefaultseppunct}\relax
\EndOfBibitem
\bibitem[Eld{\'e}n(1977)]{elden1977algorithms}
Eld{\'e}n,~L. \emph{BIT Numer. Math.} \textbf{1977}, \emph{17}, 134--145\relax
\mciteBstWouldAddEndPuncttrue
\mciteSetBstMidEndSepPunct{\mcitedefaultmidpunct}
{\mcitedefaultendpunct}{\mcitedefaultseppunct}\relax
\EndOfBibitem
\bibitem[Barth \latin{et~al.}(2004)Barth, Kuczera, Leimkuhler, and
  Skeel]{Barth2004}
Barth,~E.; Kuczera,~K.; Leimkuhler,~B.; Skeel,~R.~D. \emph{J. Comput. Chem.}
  \textbf{2004}, \emph{16}, 1192--1209\relax
\mciteBstWouldAddEndPuncttrue
\mciteSetBstMidEndSepPunct{\mcitedefaultmidpunct}
{\mcitedefaultendpunct}{\mcitedefaultseppunct}\relax
\EndOfBibitem
\bibitem[Davis(2006)]{Timothy2006}
Davis,~T.~A. \emph{Direct Methods for Sparse Linear Systems}; SIAM, 2006\relax
\mciteBstWouldAddEndPuncttrue
\mciteSetBstMidEndSepPunct{\mcitedefaultmidpunct}
{\mcitedefaultendpunct}{\mcitedefaultseppunct}\relax
\EndOfBibitem
\bibitem[Norris \latin{et~al.}(1991)Norris, Anderson, and Baker]{Norris1991}
Norris,~G.~E.; Anderson,~B.~F.; Baker,~E.~N. \emph{Acta Crystallogr. Sect. B}
  \textbf{1991}, \emph{47}, 998--1004\relax
\mciteBstWouldAddEndPuncttrue
\mciteSetBstMidEndSepPunct{\mcitedefaultmidpunct}
{\mcitedefaultendpunct}{\mcitedefaultseppunct}\relax
\EndOfBibitem
\bibitem[Forum(1994)]{Forum:1994:MMI:898758}
Forum,~M.~P. \emph{MPI: A Message-Passing Interface Standard}; 1994\relax
\mciteBstWouldAddEndPuncttrue
\mciteSetBstMidEndSepPunct{\mcitedefaultmidpunct}
{\mcitedefaultendpunct}{\mcitedefaultseppunct}\relax
\EndOfBibitem
\bibitem[Stein \latin{et~al.}(1994)Stein, Boodhoo, Armstrong, Cockle, Klein,
  and Read]{1PRT}
Stein,~P.~E.; Boodhoo,~A.; Armstrong,~G.~D.; Cockle,~S.~A.; Klein,~M.~H.;
  Read,~R.~J. \emph{Structure} \textbf{1994}, \emph{2}, 45--57\relax
\mciteBstWouldAddEndPuncttrue
\mciteSetBstMidEndSepPunct{\mcitedefaultmidpunct}
{\mcitedefaultendpunct}{\mcitedefaultseppunct}\relax
\EndOfBibitem
\bibitem[Speir \latin{et~al.}(1995)Speir, Munshi, Wang, Baker, and
  Johnson]{CCMV}
Speir,~J.~A.; Munshi,~S.; Wang,~G.; Baker,~T.~S.; Johnson,~J.~E.
  \emph{Structure} \textbf{1995}, \emph{3}, 63--78\relax
\mciteBstWouldAddEndPuncttrue
\mciteSetBstMidEndSepPunct{\mcitedefaultmidpunct}
{\mcitedefaultendpunct}{\mcitedefaultseppunct}\relax
\EndOfBibitem
\end{mcitethebibliography}

\begin{tocentry}
\center
\includegraphics[scale=0.125]{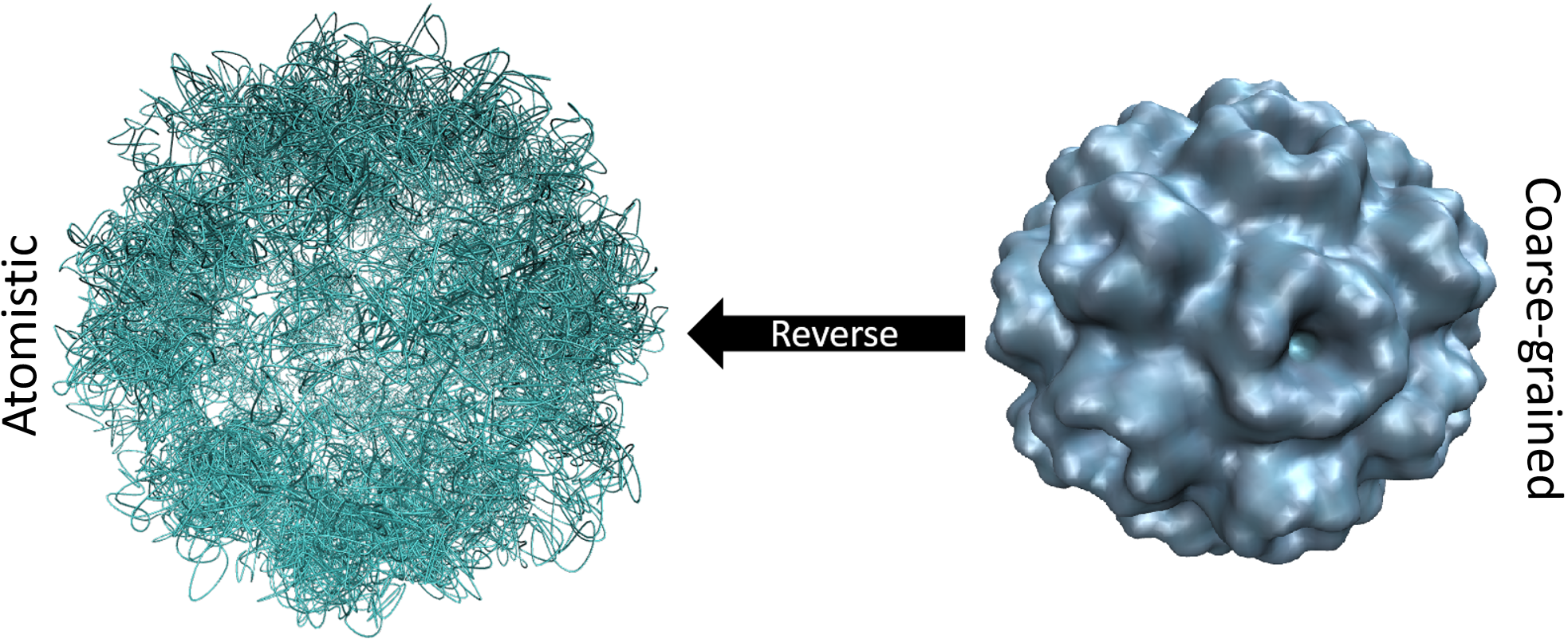}
\end{tocentry}

\end{document}